# TGL-Lambda: An implementation of TrapGrid to estimate trap attractiveness from heterogeneous field data

Ben Scalero & Nicholas C. Manoukis

## 1 Introduction

Invasive insect pests are a serious threat to agriculture around the world. Numerous costly programs are in operation today to mitigate this problem. Some research to improve these programs focuses on trap networks, which can be crucial for pest detection, control, delimitation, and eradication programs (El-Sayed). When available for the target pest insects, traps are baited with attractive lures. Various types of lures may be used for these trap networks, but not all types of lures are made equal, with some being more attractive to target pests than others. While the attractiveness of lures is an important element of the aforementioned agricultural programs, quantifying attraction can be complicated. One approach, which we build on here, can be found in Manoukis et al. (2014). In that article, the attractiveness of a lure is represented as "$\lambda$" in the hyperbolic secant function, which is used to relate the distance of an insect from a trap to the probability that it ends up being captured in the trap. It points out, for simplicity, that the reciprocal $1/\lambda$ is equal to the distance at which there is approximately a 65% probability of an insect being caught by the trap, making it easier to comprehend the meaning of the value and compare between species and lure pairs. Understanding and quantifying the levels of attraction between lures and pests is very important to determine optimal trapping strategies.

This paper describes a recently developed software called "TGL-Lambda" enables quantifying lure attractiveness under a variety of field capture scenarios including mixed lure/trap combinations. TGL-Lambda delivers a flexible approach to simultaneously estimating the $\lambda$ value for multiple trap types, accommodating a common situation in "Mark-release-recapture" (MRR) experiments in the field. Specifically, where researchers release a known number of marked insects in a field and count how many are recaptured in two to five trap and lure types, and the trap and release locations are known, TGL-Lambda can be used to estimate the attractiveness ($\lambda$) of each of the trap types.

## 2 About TGL-Lambda

### 2.1 Introducing TGL-Lambda

TGL-Lambda is a Python software that was made with the purpose of estimating levels of attractiveness for lures based on field mark-release-recapture data. TGL-Lambda includes the java software "TrapGrid" by Manoukis et al. in 2014 and uses it to test numerical hypotheses of possible combinations of $\lambda$ for each trap type, iterating to approach the observed capture probability. TGL-Lambda uses a sampling algorithm to test combinations of $\lambda$ (more details in section 2.3).

Like TrapGrid, TGL-Lambda allows the user to include the number of traps, the locations of these traps, and other parameters involved in the experiment, such as the number of flies and their diffusion coefficient – the amount of meters$^2$ per day that the insects may travel. It passes these values to executions of TrapGrid and collects resulting escape probabilities. However, a key difference is that TGL-Lambda does not need lambda values for traps in the trap location file; rather, classes of traps are indicated (e.g. "type 1", "type 2", etc.). TGL-Lambda requires a "desired escape probability". This is the value that the user obtained and calculated from their own experiment or perhaps a target sensitivity when designing a grid with traps of unknown attraction, and TGL-Lambda will attempt to find possible λs that could have resulted in this value. While TrapGrid takes in traps with known levels of attractiveness and returns an average escape probability, TGL-Lambda takes in traps with an average escape probability and returns possible levels of attractiveness.

TGL-Lambda is our first addition to a new collection of trap network software called "TG Laboratory". TrapGrid is the foundation for TG Laboratory, which will expand around it to address specific experimental or programmatic situations. TGL-Lambda is the first tool in the TG Laboratory suite. Our plan is to have more programs like this that will serve various purposes, such as delimitation or solving for other unknown variables. More about TG Laboratory will be discussed in Chapter 6 of this paper.

## 2.2 Application of TGL-Lambda

Information on communication distances in natural systems and the ranges of action of attractant-baited traps has been sought after by entomologists worldwide for quite some time (Schlyter 1992). A common but laborious method for obtaining this information is by setting up a Mark-Release-Recapture (MRR) experiment, which involves the placement of traps and the release of a known number of marked insects then measuring the number that are recaptured in each trap by checking for the marking (see Yao et al., 2022 for examples). Often, more than one trap types may be used, with multiple of each type rotated in space to avoid positional bias. From these experiments, the types of lures and the recapture rates can be recorded, allowing a quantitative estimation of trap attraction.

TrapGrid implements a mathematical formalization of trap attraction, utilizing the parameter λ as its trap attraction parameter. To our knowledge, there have been only two published studies of field experiments that were designed with the purpose of estimating λ to date (Manoukis and Gayle, 2015; Manoukis et al., 2015). TrapGrid was initially created to allow quantification of a particular trap network's sensitivity, but without parameters (particularly attraction), these estimates can't be obtained (Manoukis 2023). It would be helpful to use experimental data from research that may not have been focused on estimating λ to do so anyway.

MMR experiments can involve multiple lures, each with different levels of attractiveness. Solving for a single value of λ in an experiment given the traps, the recapture rate, and the mathematical procedure is straightforward, but this changes as you add more λs. While theoretically TrapGrid can be used to estimate trap attraction via trial-and-error, this would be cumbersome or impractical as you increase the number of different trap types being deployed simultaneously. The user will have to run TrapGrid for numerous trials, and the adjustments to be made would be tougher to figure out. For instance, how would the user know which λ to increase or decrease? Another problem with a trial-and-error approach is that there could possibly be multiple answers to what these values of λ are, as increasing one λ by a certain amount and decreasing another by a corresponding amount could potentially result in the same recapture rate.

TGL-Lambda was developed to ease the process of estimating lure attractiveness from completed MMR experiments, especially for those with multiple types of lures involved. TGL-Lambda can test these types of experiments, such as delimitation, with many different randomly generated combinations of λs to see which combinations will produce the best results.

## 2.3 Creating TGL-Lambda

TGL-Lambda can be thought of as an *extension* of TrapGrid. TGL-Lambda was created to expand on one specific potential application of TrapGrid: estimating λ. It executes a set of TrapGrid simulations after creating a list containing many of the possible λ combinations, but the number of λ combinations may be quite large. Technically, there is an infinite number of possible combinations, as the value of λ could be any positive number. Because of this, TGL-Lambda only takes λ values between 0 and 1, since any value greater than 1 would result in a $1/λ$ less than 1, which we consider the lower limit for a trap that can be considered attractive to the target insect. Even with this simplification, there are still an infinite number of values between 0 and 1.

Our approach is to instead split this range into intervals, then only select one value from each interval. By default, TGL-Lambda splits the (0, 1) range into 20 intervals, each interval only spanning 0.05. This allows many values to be tested and even if it doesn't provide an extremely accurate result, it will still give the user a solid idea of where the solution may be. However, TGL-Lambda allows any number of different lures to be input (we recommend keeping it below 5 for practical reasons like avoiding local maxima and ensuring sufficient parameter space sampling). Even if we narrow each λ into just 20 possibilities, the number of total combinations is $20^n$, where n is the number of different λs. This will result in an incredibly long list for TGL-Lambda to go process, resulting in a tremendous amount of runtime. To solve this, TGL-Lambda implements a Latin Hypercube Sampling class.

Latin Hypercube Sampling (LHS) is a statistical method used to generate random samples of parameter values (McKay). It efficiently samples from a multidimensional space while ensuring adequate coverage across all dimensions. Each dimension space, which represents a variable, is cut into n sections where n is the number of sampling points and then only one point is put in each section (SMT 2.5.0 documentation). In Figure 1 by ResearchGate below, LHS divides the parameter space (0, 1) into 5 equally spaced intervals for both dimensions. Then, within each interval, a random sample is selected. Unlike traditional random sampling, LHS ensures that each interval only contains a single sample along each dimension. Notice that for both the X and Y dimensions, going through any one of their intervals will only yield one sample. Not only that, but the sample is randomly placed within that interval. There are 5 intervals, so there will be only 5 samples, despite however many dimensions there are. By using LHS, TGL-Lambda is able to narrow down its number of combinations to only 20, by default, regardless of the number of different lures the user wants to find the λ for, while still covering the entire parameter space (0, 1).

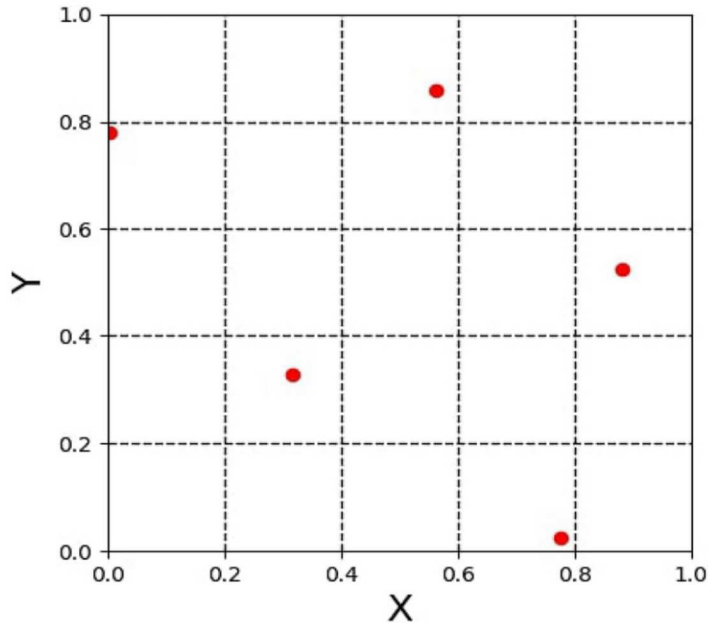

**Figure 1** An example of Latin Hypercube Sampling. Each dimension X and Y are split into 5 equal intervals of length 0.2, and one sample is randomly chosen from only one interval across both dimensions.

"Desired escape probability" is the most important parameter for TGL-Lambda and is required. The purpose of it is to tell TGL-Lambda what it is searching for. TGL-Lambda will use LHS and run TrapGrid numerous times, coming up with a large number of results. It will then take the results, which are the average cumulative escape probability from its set of λ on the final day of each simulation, and sort these by their closeness to the desired escape probability. Now, LHS makes obtaining the λs that result close to this value more realistic, but it does hinder the accuracy since there will be many values and combinations that won't be tested, and this issue only gets worse with more dimensions. This can lead to results that are not satisfactory to the user.

To solve this problem, TGL-Lambda includes an optional parameter called "tolerance". This parameter represents the maximum absolute distance that the user is willing to accept between the desired escape probability and the closest result generated by TGL-Lambda. Though the user can already identify if the best result was too far away from what was wanted, they might not know the next steps to obtain a more accurate result. Tolerance was added to allow the program to recognize this too, and it will then generate a recommendation in the form of suggested upper and lower bounds, which represent the highest and lowest values, respectively, that LHS will generate for each dimension. "Lower bounds" is set to 0 and "upper bounds" is set to 1 by default for every sample. The tolerance, bounds, and "number of lambda sets" parameters were incorporated into this software to provide the user with an easy and clear way to improve accuracy towards their desired escape probability. Further detail on these parameters and the methods that take advantage of them are available in Chapter 3.3. Lastly, an important feature of TGL-Lambda is that it contains a GUI – a graphical user interface. Many specialized packages may lack this- TrapGrid is an example as it is used on the command line. TGL-Lamba's GUI helps provide a clear way for the user to execute the program, which we demonstrate below.

# 3 How to Install and Use TGL-Lambda

## 3.1 Installing TGL-Lambda and the necessary modules

As of 2024, TGL-Lambda has been made public domain and its package is available at https://github.com/benscalero/TG-Laboratory/tree/main/TGL-Lambda.

If you would like to download all TG-Laboratory packages, you can open the TG-Laboratory directory as a whole, click the green "Code" button on the right, and download the zip files. If you only want TGL-Lambda, you can go to https://download-directory.github.io/. Here, you can paste the TGL-Lambda GitHub URL and the site will download the zip files for you. You will want to unzip/extract these files and move them to a folder that is easy to get through via the command-line interface (CLI) on your computer – this will be Command Prompt on Windows or Terminal on Apple. Next, you will need to make sure you have the necessary modules to run TGL-Lambda. If your computer already has up-to-date versions of the "numpy", "scipy", and "shapely" modules, you can skip this part. If not or if you're unsure, simply open your computer's CLI and navigate to the directory that has the unzipped TGL-Lambda files. Here, type the following into the command line:

```
py -m pip install -r requirements.txt
```

This will begin the downloading process if it is needed on your computer. When this is finished, you are ready to run TGL-Lambda.

## 3.2 Running TGL-Lambda and applying its parameters

Once you have TGL-Lambda and all of its needed modules, you are now ready to run the program. To start, open the CLI on your computer and navigate to the directory that is containing the program. Once there, you can run TGL-Lambda by typing the following into the command line:

```
py TGL-Lambda.py
```

This should open up a graphical user interface (GUI) for the user to input parameters before running the simulations. The parameters are split into 3 sections: Experiment Info, Insect Info, and Results Info. This chapter goes over the different parameters and what they mean to TGL-Lambda, but for the convenience of the user, there are "more info" buttons at the start of each section within the GUI, and the parameters' details are given there as well.

**Figure 2** TGL-Lambda GUI. After running py TGL-Lambda.py, this window should appear, displaying all of the possible parameters for the user to input before running the simulations.

The first section is Experiment Info, which consists of the "TrapGrid file", "Outbreak file", "number of days", "number of simulations", "number of flies", and "random seed" parameters. If you have used TrapGrid before, these parameters should all be familiar to you. "TrapGrid file" is a required parameter that provides TGL-Lambda with the details of the experiment's grid layout and trap details. This should be a tab-separated values (.tsv) file. The first row should give the grid dimension as x meters by y meters, written in the file as just the numerical values separated by a tab. Every following row after this represents a trap and should consist of 3 values, again all separated by tabs. The first value is the x-coordinate of the trap, the second value is the y-coordinate of the trap, and the third value is the λ parameter for the trap. In TrapGrid, this is supposed to be the known numerical value of λ, but TGL-Lambda is meant to solve for this. In order for TGL-Lambda to do so, the user must put a placeholder name for the λ. This is used to differentiate between different types of lures. For example, if the first three traps consist of one type of lure, the user can put "lam1" for their λs in the TrapGrid file, but if the next six traps consist of another type of lure, "lam2" should be put instead.

**Figure 3** Sample TrapGrid file text document. This is a tab-separated-values (*.tsv) file. The first line is the dimensions of the grid in x meters and y meters, so in this example, it is a 1000m x 1000m grid. Each subsequent line after represents a trap. The first trap shown has a (x, y) location of (250, 250), and consists of a lure with an unknown λ. As you can see from the file, the first three all share the same type of lure, while the last six share a different kind.

"Outbreak file" is an optional parameter that consists of the (x, y) locations of outbreaks. Again, this file must be tab-separated. If given, TGL-Lambda will run simulations using each of these as the outbreak location. If this file is not provided, the program will simply run simulations with randomized outbreak locations. The next parameters in the Experiment Info are much more straightforward. "Number of days" indicates how many days pass in each simulation of the experiment, "number of simulations" indicates how many simulations will occur for each set of λs, "number of flies" indicates how many flies are to be released for each simulation, and "random seed" is just an arbitrary value that the user can provide for the randomization process and allows replication between instances.

The second section is Insect Info, and this provides detail on the insects themselves. Again, these parameters are no different than the ones in TrapGrid. "Diffusion coefficient" is the first parameter, and it indicates the distance that the flies move in meters$^2$ per day. If this value is provided, ignore the rest of the parameters in this section. Otherwise, the user can provide the step size, number of steps per day, and the turn angle standard deviation of the insects. If these are provided, then the program will use a Mean Dispersal Distance model instead. However, if none of the "Insect Info" parameters are given, TGL-Lambda will use the diffusion model with a diffusion coefficient of 30.0 by default.

The third section is Results Info. These parameters are what differentiate TGL-Lambda from TrapGrid, and they control the program's output. The first parameter is "Desired Escape Probability", and it is required as it is the value that all of TGL-Lambda revolves around. This value is what the user wants the average cumulative escape probability to be after the final day of simulations. After the program runs every simulation, it will print every set of λ combinations and their corresponding final day result, sorted by their result's distance to this value. So, the result shown first will represent the λ combination that gave an escape probability closest to this value. The next parameters are optional and are focused on improving the program's accuracy. The first of these is "tolerance". This is a value greater than 0 but less than 1 that represents the maximum distance that the user wants between their desired escape probability and the closest result that the program returns. Essentially, this is a way for the program to understand whether its results end up being satisfactory to the user or not. If that distance between the desired and the top result is greater than the tolerance, the program will give the user a "recommended" lower and upper bounds of the λ values for the user to include in another run of TGL-Lambda, which should allow for more accurate results. "Lower bounds" and "upper bounds" are the next two parameters, and they represent the lowest and highest, respectively, values of λ that the program will obtain in its sampling. These should be inputted as lists, starting and ending with [] brackets, and they should contain the λ values inside, separated by a comma. If these two parameters are not provided, which is recommended for the first trial of TGL-Lambda unless you have an idea of the range λ may be in, then lower bounds will be set to a list consisting of all 0s, while upper bounds will be set to a list consisting of all 1s.

**Figure 4** An example of what the "Results Info" section of the parameters can look like. In this example, LHS will generate 30 sets. Throughout these sets, the samples for the first λ parameter will only be between 0.15 and 0.9, the second will be between 0.2 and 0.8, and the third one will be between 0.4 and 0.7. The results will be sorted by their closeness to 0.5, and if the closest one is further than 0.01 from this value, a recommended adjustment to the bounds will be given.

The final parameter focused on accuracy is "number of lambda sets", which is the value of how many different sets of λ for TGL-Lambda to create, using Latin Hypercube Sampling, for testing. By default, this value is 20, meaning that it uses LHS where the parameter space (0, 1) is split into 20 intervals that span 0.05 each. If the user were to set this value to 50, for example, it will split the space into 50 intervals that span 0.02 each, resulting in more combinations of λs to be tested, therefore improving accuracy. However, keep in mind that the greater this value, the longer the runtime as well, as TGL-Lambda will be testing more. Finally, the last parameter of TGL-Lambda is "output file". This is an optional .txt file for the program to put its results. This way, the user can see the results in a more readable display, and it also allows these results to be saved and stored.

After all the parameters are set, you can click the "Run Simulation" button on the bottom and the simulations will begin. Even if you gave an output file – which is highly recommended – you will still see some output on the CLI, which is just there for you to see the program's progress. It will keep the user updated on the day it is simulating, which simulation it is on, and for what set of λs out of however many it will run. When completed, the results will be either in the CLI or in the given output file, and the program will automatically close.

## 3.3 Methods to improve accuracy

TGL-Lambda has several methods and parameters that can improve the accuracy of the results. We've already gone through two things that can improve accuracy. One of these was that you can simply increase the number of λ sets for LHS to make. The other was under the condition that the top result was outside of the tolerance, for which you would take the recommended bounds given and run TGL-Lambda again with these lower and upper bounds. The following instance would provide a more accurate result than the previous one. If the new result were to still outside of the tolerance, this procedure can be done again, as many times as needed, until an adequate result is achieved. This finds a range that the λs can be in and narrows it down, pinpointing its location. However, keep in mind that when multiple λs are involved, there can be multiple solutions. For example, maybe TGL-Lambda was given two λs to solve for, and it found a solution where the first λ is a high value and the second λ is low. This is not to say that there isn't a case where the first λ is lower and the second λ is higher, resulting in the same escape probability. This just means that the λ values TGL-Lambda used happened to be closest to that first solution, and it narrowed in on that set.

To help avoid this problem, you can run multiple instances of TGL-Lambda – at the same time if you want – and split the λ bounds by sections. For example, say you are running TGL-Lambda with two unknown λs. You can run the first instance with a lower bound of [0, 0] and upper bound of [0.5, 1]. The second instance could be [0.5, 0] and [1, 1]. The third instance could be [0, 0] and [1, 0.5], and the last one could be [0, 0.5] and [1, 1]. As you can see, the first two instances will be mainly searching for the first λ, splitting it up into two part to find if there's a low value solution and if there's a higher value solution. The last two instances will do the same but for the second λ. This is a great way to see if the general location of possibly more than one solution, and then you can begin pinpointing them individually.

# 4 Practicum: A Step-By-Step Example

## 4.1 Background Information

Let's say we already have the results of an MRR experiment that we completed. This experiment consisted of a 200m x 600m grid, 60 traps, 90 outbreak locations, and about 40,000 flies. The traps are split into four different groups, each group containing a different lure. These groups are labeled 'A', 'B', 'C', and 'D'. What we already know about these lures are that they are very similar, but slightly vary in attractiveness. As of right now, their estimated λs are all 0.07.

After collecting the results of the experiment, the proportion of escaped flies is 0.40. Although the different types of traps caught about the same number of flies, it is important to note which ones did better than others. Group 'A' caught the most, group 'C' caught the least, and group 'B' and 'D' caught nearly the same amount. So, we are looking for λs where group 'A' has the smallest value, group 'B' and 'D' are equally greater (but not by much), and group 'C' is the greatest.

## 4.2 Applying TGL-Lambda

Our recorded proportion of escaped flies is 0.4, so this will be the desired escape probability in our instances of TGL-Lambda. Since we already have the data, what's left before running the software is to create the TrapGrid file and the Outbreak file. Remember that this involves the known grid size and coordinates from the experiment. The first ten lines of each file are given below in Figure 6.

**Figure 5** The first ten lines of the TrapGrid file created for this experiment. Each trap is denoted by a (x, y) location and its λ parameter, depending on which type of lure was used for the trap.

After creating these files, we are ready to run TGL-Lambda. In the command prompt, go to the directory that contains TGL-Lambda and run **py TGL-Lambda.py**. This will launch the program and opens the window where you input the data. First, we select our TrapGrid file and Outbreak file. We complete the "Experiment Info" portion by putting in number of days, number of simulations, and number of flies. The experiment consisted of seven days and around 40,000 flies, which comes out to about 450 flies per release point. As for the number of simulations, no input is needed since an outbreak file was provided and the program will automatically run one simulation per release point given by the file. Moving on to the "Flies Info", the diffusion coefficient we will use for these flies is 5000 $m^2$/day.

Now, it is only a matter of filling out the Results portion. Our desired escape probability is 0.4, and we'll make the tolerance 0.01. The estimated λ for the lures are all 0.07, but running this in TrapGrid gives an escape probability of about 0.33. Since we want the escape probability to be a bit higher and the values of the λs are likely not that far apart, we can guess that the λs would be between 0.07 and 0.075. So, since we have four different λs, the lower bounds input is "[0.07, 0.07, 0.07, 0.07]", and for upper bounds, it is "[0.075, 0.075, 0.075, 0.075]". We will also run 100 sets of λs, which took a long time to run but provided a wide range of results. Then, I created a .txt file for the output, and ran the simulations.

**Figure 6** Image of the TGL-Lambda GUI filled out for this experiment example. This is the first instance of TGL-Lambda out of three total. "Lambda lower bounds" and "Lambda upper bounds" were the only parameters adjusted for the next two instances.

## 4.3 Reviewing the results

This experiment example was done on a laptop with an Intel(R) Core(TM) i7-7700HQ CPU @ 2.80GHz processor and 16GB of RAM. Each execution of TGL-Lambda for this experiment ran in the background for about 5-6 hours. This long duration is due to the large number of flies, traps, release points, and λ sets. Once the command-line says that the simulations were completed, we can open the file selected for output. Scrolling down to the "Final Results" section, the program gave a recommendation, which means that none of its results were within the tolerance given and another instance of TGL-Lambda should be run. In Figure 7 below, it can be seen that the recommended upper bounds that it gave was a list of ones. This means that out of all the sets of λs that it ran, none of them resulted in an escape probability higher than the desired one. Taking a look at the closest result also shows that even though all λs were close to 0.075, the resulting escape probability was still not high enough.

**Figure 7** The output file from the example execution of TGL-Lambda. The first escape probability result is the closest to our desired escape probability of 0.4, and since it is too far ( > 0.01 away), the output gives recommended bounds for running the next instance.

Using this knowledge, we can make the lower bound a list of 0.075 and the upper bound 0.08. While the recommended ones are sufficient, it is safe to assume that the λs must be greater than 0.075, and since the closest result was still close to being within the tolerance, they must be close to this value.

After running TGL-Lambda again with the same inputs except for the different bounds and output file, the results looked similar to the first one. Although the closest escape probability was closer, its set of λs were all nearly 0.08 and the recommended upper bounds was once again a list of ones. So, the same process was repeated, but instead of a λ range of 0.075-0.08, it was a range of 0.08-0.085. Finally, no new recommendation was given. In fact, multiple results were within the tolerance, any of which can be our answer since a simulation has a small amount of randomness to where the flies go. For this experiment, it was recorded that the 'A' traps caught more than the others, which means it would have the lowest λ. It was also noted that the 'C' traps caught the least and the other two were about the same. Looking at the results, there is a set of λs that fits this description and results in an average escape probability within the tolerance. So, the program's answer for the values of λs is:

[0.080, 0.082 0.085, 0.082]

It is important to remember that this is still an approximation. To further test accuracy, we can take these values of λs and put them into our TrapGrid file and then run TrapGrid, with a large number of simulations to see if it still results in our desired escape probability. TrapGrid showed that these λs result in an escape probability that is slightly larger than what we want. This approximation is still quite accurate, but if a more accurate answer is wanted, TGLambda can be ran again with a tighter range, less sets, and more simulations per set. Another thing to keep in mind is that this is under our assumption that the λs are very close, which is why we were able to use 0.005 ranges. It might be smart to run this again but with a wider range for λ, such as 0.075-0.085, to see if this is still about the result we get.

## 5 Discussion

In the example above, Lure A is the most attractive with an estimated λ = 0.080, which means that there is a 65% probability of capture when this lure is 1/0.080 = 12.50 meters away. For lures B and D, this distance is 1/0.082 = 12.12 meters, and for lure C, it is 1/0.085 = 11.76 meters. Now that the λs are quantified, further experimentation can be done with this knowledge in hand. These numbers give a numerical idea of what to expect from each lure, but there is more that you can do. You can plug the λ values right into your TrapGrid file and test out different grids on TrapGrid. You can also replace all λs with the best value obtained and then run TrapGrid to see what the experiment may have looked like if you had only used the strongest lure. A comparison between trap/lure combination scenarios might also be used to estimate costs of alternative trap network plans and develop cost-benefit analyses.

TGL-Lambda can be improved. While the program gives an assessment of the attractiveness of the lures, the quality of running the program itself can still be better. There are plans to add changes to the TGL-Lambda GitHub publication, which include being able to input both known and unknown λs, having the option to choose between LHS or Normal (full factorial) Sampling, and more. Be sure to check the ReadMe file on GitHub occasionally to see if you have the most up-to-date version.

## 6 Conclusion

Goals for experiments with trap networks vary. TrapGrid and TGL-Lambda are useful tools for quantifying capture and attraction. However, what if we wanted to estimate diffusion coefficients or to optimize the cost of the experiment? Maybe we could use a program that continuously increases the number of traps until adding more doesn't provide a significant enough benefit. There are also ways to ease the use of these programs, such as a grid generator that creates a TrapGrid file based on trap density per square mile for delimitation purposes. We hope to add additional useful tools through the expansion of TG Laboratory.

# 7 Acknowledgements

I would like to thank Blue for her support and encouragement when I first thought of this idea, and thank you to my team, Lotte, Nic, Hyoseok, and Kristen, at the USDA. Thank you to Blue and Jo for reviewing this paper. Mention of trade names or commercial products in this publication is solely for the purpose of providing specific information and does not imply recommendation or endorsement by the US Department of Agriculture. The USDA is an equal opportunity provider and employer.